\documentclass[12pt,preprint]{aastex}
\usepackage{emulateapj5}

\def\gsim { \lower .75ex \hbox{$\sim$} \llap{\raise .27ex \hbox{$>$}} }
\def\lsim { \lower .75ex \hbox{$\sim$} \llap{\raise .27ex \hbox{$<$}} }

\shorttitle{Torques and Orientation of Disk Galaxies}
\shortauthors{Navarro, Abadi \& Steinmetz}

\begin{document}


\title{Tidal Torques and the Orientation of Nearby Disk Galaxies}

\author{Julio F. Navarro\altaffilmark{1,2}, Mario G. Abadi\altaffilmark{1,3}, and Matthias Steinmetz\altaffilmark{4}}

\begin{abstract}
We use numerical simulations to investigate the orientation of the angular
momentum axis of disk galaxies relative to their surrounding large scale
structure. We find that this is closely related to the spatial configuration at
turnaround of the material destined to form the galaxy, which is often part of a
coherent two-dimensional slab criss-crossed by filaments. The rotation axis is
found to align very well with the intermediate principal axis of the inertia
momentum tensor at this time. This orientation is approximately preserved during
the ensuing collapse, so that the rotation axis of the resulting disk ends up
lying on the plane traced by the protogalactic material at turnaround. This
suggests a tendency for disks to align themselves so that their rotation axis is
perpendicular to the minor axis of the structure defined by surrounding
matter. One example of this trend is provided by our own Galaxy, where the
Galactic plane is almost at right angles with the supergalactic plane (SGP)
drawn by nearby galaxies; indeed, the SGP latitude of the North Galactic Pole is
just 6 degrees. We have searched for a similar signature in catalogs of nearby
disk galaxies, and find a significant excess of edge-on spirals (for which the
orientation of the disk rotation axis may be determined unambiguously) highly
inclined relative to the SGP. This result supports the view that disk galaxies
acquire their angular momentum as a consequence of early tidal torques acting
during the expansion phase of the protogalactic material.
\end{abstract}

\altaffiltext{1}{Department of Physics and Astronomy, University of
Victoria, Victoria, BC V8P 1A1, Canada} 
\altaffiltext{2}{Fellow of CIAR and of the J.S.Guggenheim Memorial
Foundation}
\altaffiltext{3}{CITA National Fellow, on leave from Observatorio
Astron\'omico de C\'ordoba and CONICET, Argentina}
\altaffiltext{4}{David and Lucile Packard Fellow. Astrophysikalisches
Institut Potsdam, An der Sternwarte 16, Potsdam 14482, Germany and
Steward Observatory, University of Arizona, Tucson, AZ 85721, USA}

\section{Introduction}
\label{sec:intro}

In hierarchical models of galaxy formation, the origin of galactic angular
momentum is ascribed to tidal torques operating early on the material destined
to form a galaxy. Over the years, starting from the ideas of Stromberg (1934)
and Hoyle (1949), and quantified by the work of Peebles (1969), Doroshkevich
(1970), and White (1984), a number of important properties inherent to the
acquisition of angular momentum through tidal torques (hereafter referred to,
for short, as ``tidal torque theory'', or TTT) have been identified.

The efficiency of the torquing is low, typically endowing galactic material with
a rather small amount of net rotation. This implies that large collapse factors
within massive dark matter halos are needed to explain the
centrifugally-supported nature of galactic disks (Fall \& Efstathiou 1980).  To
leading order, galaxy spins result from the misalignment between the principal
axes of the inertia momentum tensor (${\bf I}_{ij}$) of the material being
torqued and of the ``shear'' tensor (${\it \bf T}_{ij}=-\partial^2\phi/\partial
x_i \partial x_j$) generated by external material. In the Cartesian principal
axis frame of the protogalaxy, the leading term of the torque is given by
$\tau_i = dL_i/dt \approx T_{jk}(I_{jj}-I_{kk})$ (here $i$, $j$, and $k$ are
cyclic permutations of $1$ to $3$, and $L_i$ are the cartesian components of the
angular momentum). The inertia term is maximal along the direction that
maximizes the difference between $I_{jj}$ and $I_{kk}$; i.e., the {\it
intermediate} axis of inertia. Angular momentum growth is typically linear with
time at early times (since $\Omega\approx 1$ then) and effectively ends at
turnaround, when ${\bf I}$ is maximal. In general, then, the direction of the
angular momentum will be determined by the {\it shape} of the protogalactic
material at turnaround, and is expected to align with the intermediate axis of
inertia if ${\it \bf T}$ and ${\bf I}$ are uncorrelated.

\begin{figure*}[t]
\includegraphics[height=0.35\textheight,clip]{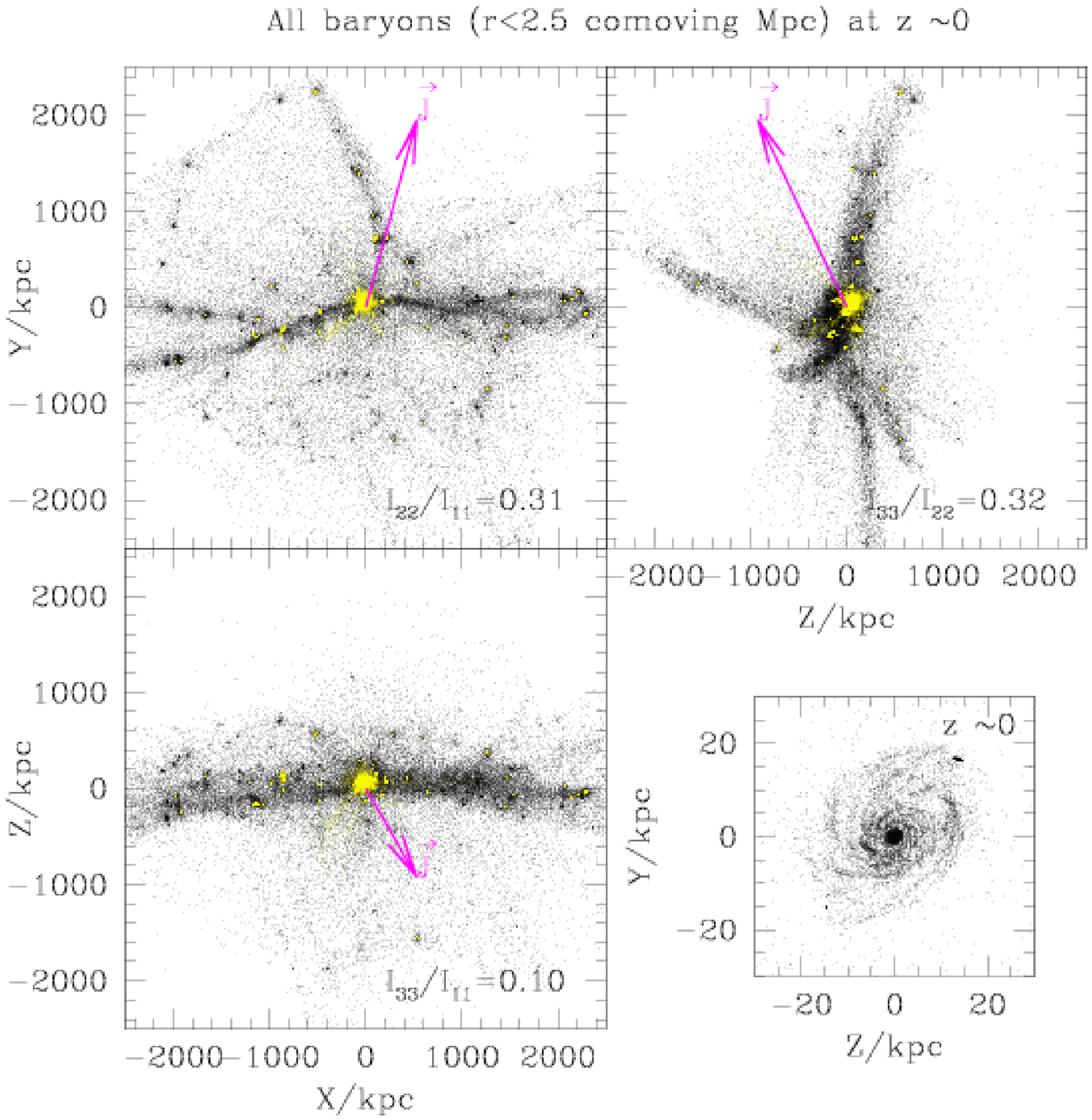}
\includegraphics[height=0.35\textheight,clip]{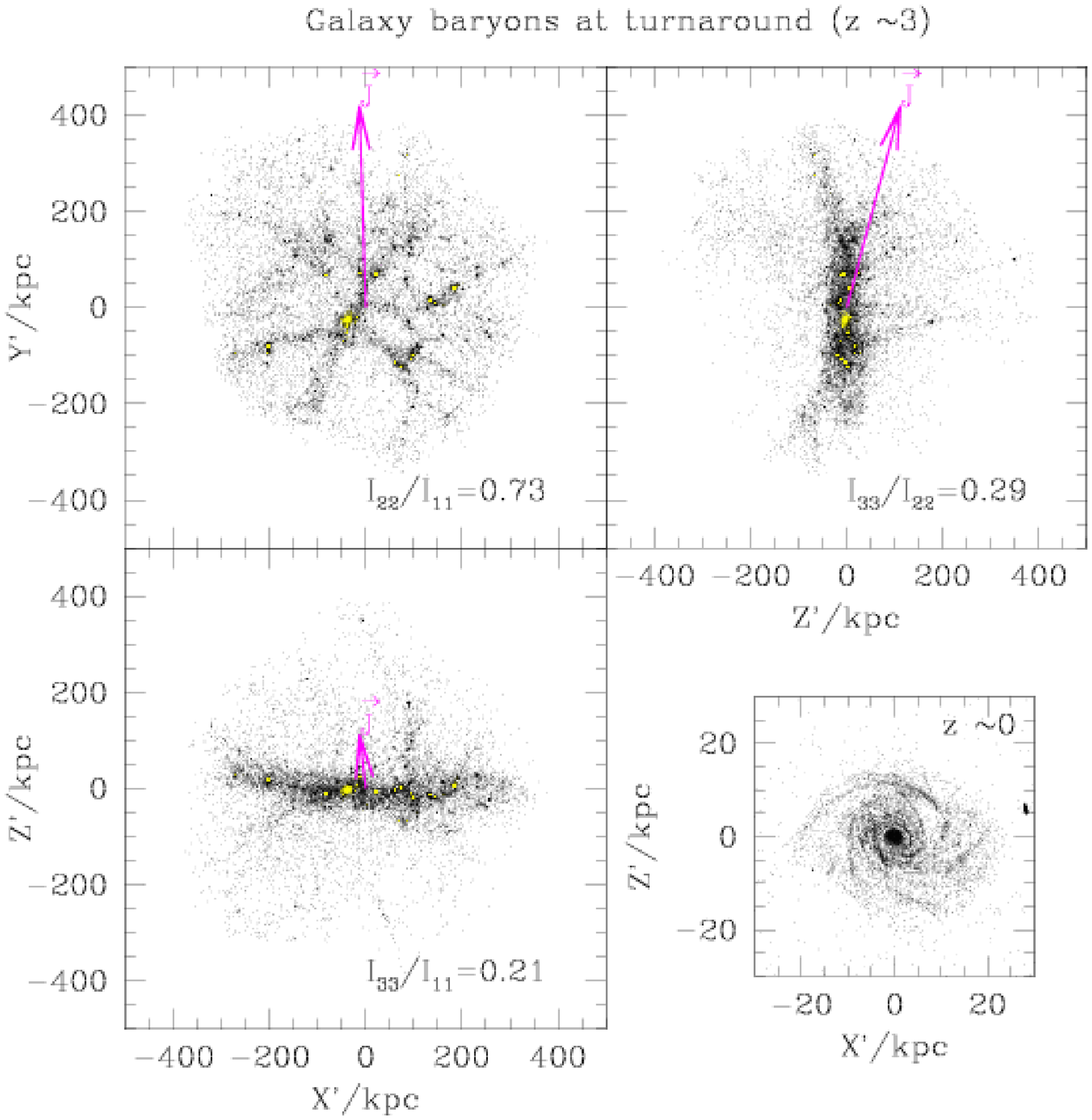}
\caption{ ({\it a-left}) Spatial distribution of all baryons in the
resimulated high-resolution region at late times. Three orthogonal projections
are shown, aligned with the principal axis of inertia of the system at that
time. Units are in physical kpc. Particles in yellow are stars; others are
gas. Labels in each panel indicate the aspect ratio of the system. Arrows
indicate the relative importance of the cartesian components of the angular
momentum. 
Bottom-right panel is a zoom of the $Y$-$Z$ projection that shows the
gaseous disk almost face-on, implying that its rotation axis lies on the plane
of the surrounding structure. ({\it b-right}) Same as (a) but for the baryons
that will collapse to form the galaxy at $z=0$, shown at turnaround in its own
principal axis frame. 
The bottom-right panel shows a zoomed-in $(X',Z')$ projection of the
same baryons at $z\sim 0$, which have collapsed to form a disk whose
rotation axis remains closely aligned to the intermediate ($Y'$) axis
at turnaround.
\label{figs:fig1}}
\end{figure*}

Many of these expectations have been confirmed by a number of studies of the
angular momentum properties of dark matter halos formed in cosmological
simulations (see, e.g., Barnes \& Efstathiou 1987, as well as the more recent
work of Porciani, Dekel \& Hoffman 2002a,b, and references therein). The latter
authors point out that ${\bf T}$ and ${\bf I}$ are, in fact, highly correlated,
and conclude that, relative to the principal axes of ${\bf T}$, the most
significant trend is for the spin to be perpendicular to the direction of
maximum compression (i.e., that of the minimum eigenvector of ${\bf
T}$). Because ${\bf T}$ and ${\bf I}$ are correlated, the direction of maximum
compresion roughly coincides with the direction of the minor axis of inertia at
the time of turnaround. A solid prediction of TTT is, therefore, that galaxy
spins should be nearly perpendicular to the minor axis of the collapsing
material at late times.

The main shortcoming of this work is that it applies mainly to dark
halos, rather than to the luminous (observable) component of
galaxies. Simulations that include the presence of a dissipative
gaseous component have highlighted the possibility that large losses
of angular momentum may accompany the collapse of the baryonic
component (Navarro \& Benz 1991, Navarro \& White 1994, Navarro \&
Steinmetz 1997).
The rather indirect mapping between the angular momentum of
dark halos and that of their baryonic components makes it difficult to assess
the success of TTT in accounting for the spin of spiral galaxies. 
The discussion of the previous paragraphs, however, suggests that TTT
may be tested by looking for residual correlations originating from
the coupling between the shear and inertia momentum tensors that
dominates the angular momentum growth in protogalaxies. One example of
this is the recent work of Lee \& Pen (2000, 2001, 2002, see also Lee,
Pen \& Seljak 2000), who attempt to reconstruct the shear field from
the observed galaxy spin fields and to uncover correlations between
the spin field and the large-scale distribution of matter.

We investigate here a related signature expected from TTT; i.e., the correlation
between the ``shape'' of protogalactic matter distribution at turnaround and the
direction of the resulting angular momentum. 
As explained above, TTT suggests a trend for spin
axes to be perpendicular to the minor axis of the sheet, so that disk galaxies
should be highly inclined relative to the plane defined by their surrounding
structure. 

In this {\it Letter} we explore whether disk galaxies assembled hierarchically
in $\Lambda$CDM numerical simulations indeed follow these expectations. In
addition, we search for a statistical signature of the predicted trend in
catalogs of nearby galaxies. 


\section{Numerical Experiments and Results}
\label{sec:numexp}
We have analyzed four N-body/gasdynamical simulations of the formation of disk
galaxies in the ``concordance'' $\Lambda$CDM cosmogony ($\Omega_0=0.3$,
$\Omega_{\Lambda}=0.7$, $h=0.65$, $\Omega_b=0.019\, h^{-2}$,
$\sigma_8=0.9$). These simulations are similar to those described in detail by
Abadi et al (2003a,b) and Meza et al (2003), where we refer the reader for full
details.  The simulations follow self-consistently the evolution of a small
region surrounding a target galaxy, excised from a large periodic box and
resimulated at higher resolution preserving the tidal fields from the whole box.
For the purposes of the discussion here all four simulations give consistent
results. We have chosen to illustrate these using the simulation presented by
Abadi et al (2003a,b), but will use the others to verify the general
applicability of our results.

Figure 1a shows, at late times ($z\sim 0$), the baryons within $\sim 2.5$ Mpc of
the target galaxy. The three different projections in the figure have been
chosen so as to coincide with the principal axes of the inertia momentum
tensor. This figure shows the high degree of anisotropy that characterizes the
non-linear evolution of structures in a $\Lambda$CDM universe. What started off
as a roughly spherical region $\sim 5$ comoving Mpc across develops into a
coherent triaxial structure that surrounds the target galaxy (located at the
center of the panels in Figure 1a). The minor axis of the structure, in
particular, is very stable across the structure, which may thus be described as
a ``sheet'' criss-crossed by filaments. The ratios of the inertia tensor
eigenvectors are given in Figure 1a, and are not dissimilar to the mean of all
four systems: $\langle I_{22}/I_{11}\rangle =0.419$ (with dispersion
$\sigma=0.360$), $\langle I_{33}/I_{11} \rangle =0.174$ ($\sigma=0.157$) and
$\langle I_{33}/I_{22} \rangle =0.408$ ($\sigma=0.059$).

\begin{figure*}[t]
\includegraphics[height=0.35\textheight,clip]{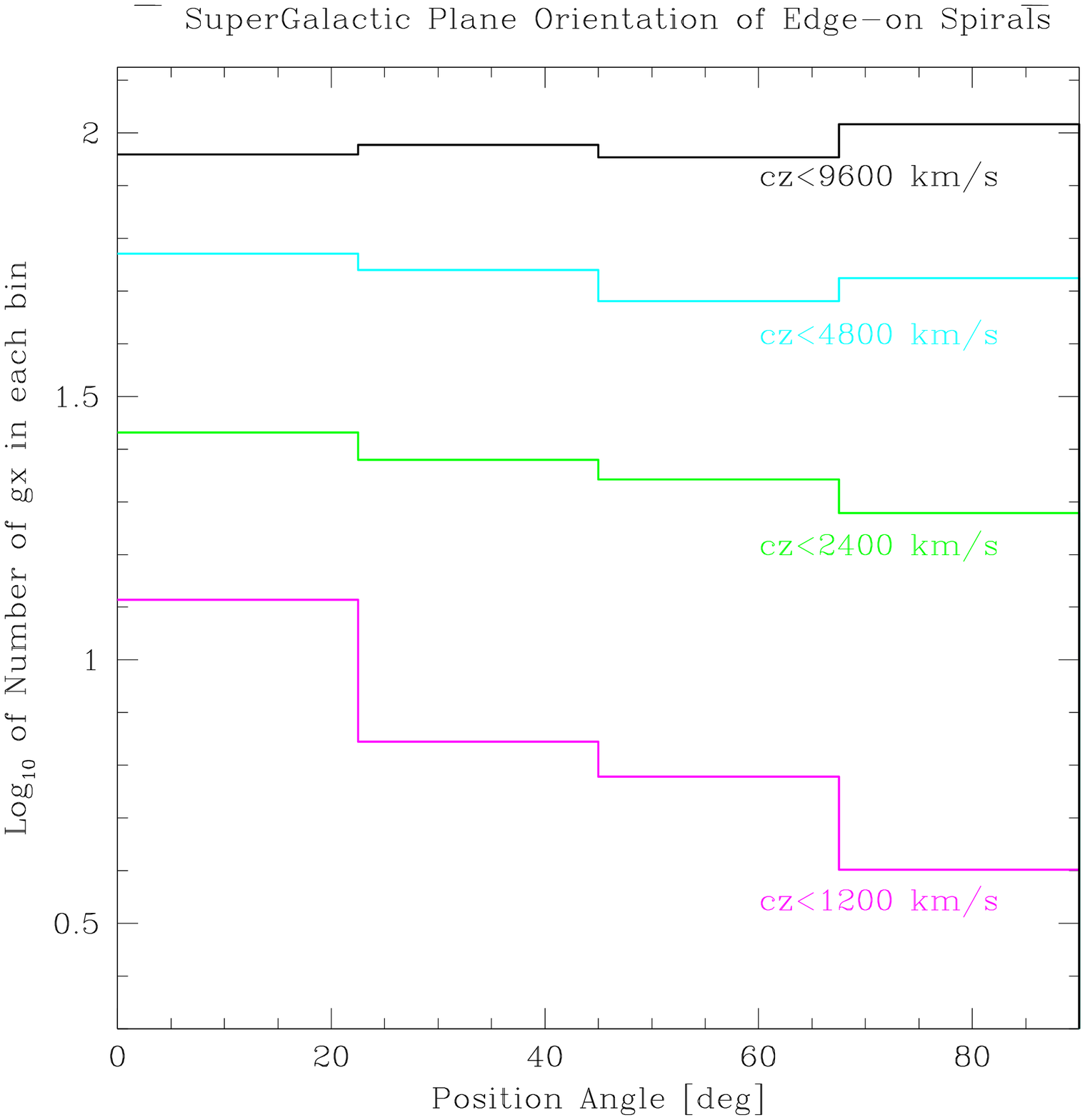}
\hspace*{+.5cm}
\includegraphics[height=0.35\textheight,clip]{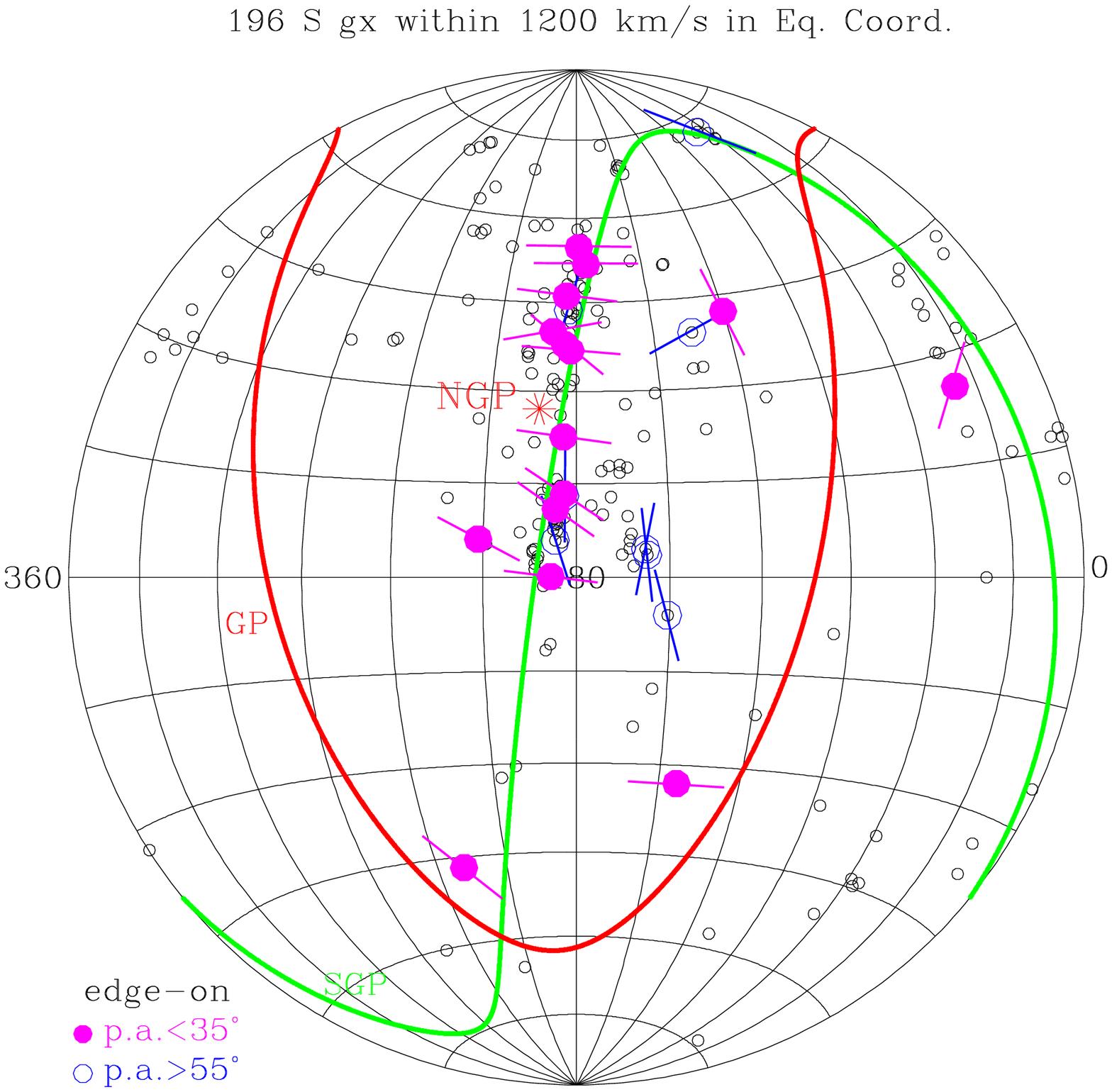}
\caption{({\it a- left}) Histogram of supergalactic position angles of
edge-on ($b/a<0.175$) spirals in the Principal Galaxy Catalog (PGC, Paturel et
al 1997) selected within various recession velocity limits, as labelled. A
position angle of $0^{\circ}$ means that the galaxy's plane is {\it
perpendicular} to the SGP; $90^{\circ}$ means that it is {\it parallel} to the
SGP. There is a well-defined excess of spirals perpendicular to the SGP in the
vicinity of the Milky Way. 
({\it b-right}) Aitoff equatorial projection of all spirals
within $1200$ km s$^{-1}$ in the PGC. The U-shaped thick curve is the Galactic
plane (GP); the S-shaped curve is the SGP. The projected major axis of edge-on
spirals is shown for galaxies with position angles smaller than $35^{\circ}$
(filled circles) and greater than $55^{\circ}$ (open circles). 
\label{figs:fig2}}
\end{figure*}


The arrows in the panels of Figure 1a indicate the relative magnitude of the
cartesian components of the angular momentum of the sheet, and illustrate the
TTT predicted trend (see \S1) for spins to be approximately perpendicular to the
direction of maximum compression or, roughly, to that of the minor axis ($Z$)
during the non-linear stages of the evolution. Indeed, the whole sheet of
material is slowly spinning around the {\it intermediate} principal axis of
inertia (i.e., $|L_Y| \gg |L_X|$, $|L_Z|$). This is a generic result of our four
simulations; quantitatively, we find that the angular momentum of the
high-resolution region surrounding the target galaxies to be oriented such that
$\langle |L_X|/|{\vec L}| \rangle =0.255$ ($\sigma=0.104$), $\langle
|L_Y|/|{\vec L}| \rangle =0.860$ ($\sigma=0.103$), and $\langle |L_Z|/|{\vec
L}|\rangle =0.351$ ($\sigma=0.275$).

This two-dimensional structure is present since early times and, at $z\sim 3$,
its minor axis is approximately parallel to that of the material that will
collapse to form the target galaxy at $z=0$ (i.e., the ``protogalaxy'', see
Figure 1b, as well as Sugerman, Summers \& Kamionkowski 2000 for a similar
analysis). The shape of the protogalactic material at turnaround
\footnote{ We define as turnaround the time when the moment of inertia of the protogalaxy
attains its maximum.}
is indicated in the the panels of Figure 1b; for the four simulations we find
$\langle I_{22}/I_{11} \rangle =0.652$ (with dispersion $\sigma=0.0.247$),
$\langle I_{33}/I_{11}\rangle =0.265$ ($\sigma=0.095$) and $\langle
I_{33}/I_{22} \rangle=0.429$ ($\sigma=0.132$). The angular momentum of the
protogalaxy at $z\sim 3$ shows similar alignment properties to that of its
surrounding material at $z\sim 0$; in particular, the close alignment with the
intermediate inertia axis at turnaround is well defined in all four simulations:
$ \langle |L_{X'}|/|{\vec L}| \rangle=0.055$ ($\sigma=0.031$), $\langle
|L_{Y'}|/|{\vec L}|\rangle =0.832$ ($\sigma=0.260$), and $\langle
|L_{Z'}|/|{\vec L}| \rangle =0.421$ ($\sigma=0.319$).

The angular momentum direction of the baryons at turnaround is approximately
preserved by the disk component at $z=0$ and, as a result, the plane of the disk
is highly inclined relative to the plane defined by its surrounding large-scale
structure.  This is illustrated in the bottom right panels of Figures 1a and 1b,
where the disk at $z=0$ is seen approximately face-on in projections that
contain either the minor axis of the surrounding structure at $z\sim 0$ or that
of the protogalaxy at turnaround. This indicates that the shape of the
protogalactic material during the expansion phase effectively determines the
orientation of the spin axis of the resulting galaxy. Since these anisotropies
correlate across a wide range of scales in $\Lambda$CDM, a signature of this
process might be preserved as a residual alignment between the disk and its
surroundings; the rotation axis of disk galaxies may be expected to lie {\it on}
the plane defined by its surrounding structure. We explore next whether such
alignments are indeed present in samples of disk galaxies in the nearby
Universe.

\section{Orientation of Nearby Disk Galaxies}
\label{sec:obs}

As recognized by de Vaucouleurs (1953) in the Shapley-Ames catalog, galaxies in
the vicinity of the Milky Way are arranged in a two-dimensional slab usually
referred to as the ``super galactic plane'' (SGP). This structure stands out
clearly in whole-sky catalogs of nearby galaxies; extends out to several tens of
Mpc from the Galaxy; and includes many of the major features of the local large
scale distribution of matter, such as the Virgo cluster, the Perseus-Pisces
supercluster, and the Great Attractor (Lahav et al 2000).

The plane of the Galaxy is highly inclined relative to the SGP; indeed, the two
planes are approximately perpendicular to each other, as indicated by the low
supergalactic latitude of the North Galactic Pole ($\sim 6^{\circ}$). This
situation is similar to that of the simulated disk galaxy in relation to its
surrounding structure, as discussed in \S\ref{sec:numexp}. It is therefore
tempting to regard the rather peculiar orientation of the Galactic plane
relative to the SGP as a result of early torques acting during the protogalactic
stage.

If this interpretation is correct, we would expect an excess of nearby galaxies
whose rotation axes are approximately perpendicular to the normal to the
SGP. The existence of such alignment has proven controversial; prior work has
led to claims of statistical evidence for such ``anti-alignment'' of galactic
planes relative to the SGP (see, e.g., Flin \& Godlowski 1989 and references
therein), but also to suggestions that there is no convincing evidence for
deviations from a random distribution of orientations (see, e.g., Dekel 1985)

A number of factors complicate this analysis and may explain this
disagreement. For example, the statistics of the position angles of galaxy
samples that do not discriminate properly between elliptical and disk galaxies
may be inconclusive, as the projected major axis of early type galaxies---in
contrast to those of disks---may bear little relation to the angular momentum of
the galaxy.  In addition, determining the direction of the rotation axis of a
spiral galaxy requires knowledge not only of the position angle and inclination
(readily available in most catalogs), but also of {\it how} the galaxy is
inclined on the sky, i.e., which side of the galaxy is closer to the
observer. This is known for a few well-studied spirals, such as M31
\footnote{The Andromeda galaxy is only mildly inclined relative to the
SGP---the supergalactic latitude of M31's pole is $\sim 56$ degrees.}
, but it is generally unavailable for most galaxies. The uncertainty introduced
by this ambiguity depends on inclination: it is maximal for galaxies inclined by
45 degrees, but is negligible for galaxies seen either face-on or edge-on.  In
this {\it Letter} we restrict our analysis to edge-on spirals and search for a
possible correlation between the orientation of galaxy disks and the SGP. We
explore ways of circumventing the aforementioned ambiguity and extend this
analysis to all nearby spirals in a forthcoming paper.

Figure~\ref{figs:fig2}a shows the distribution of position angles
(in the supergalactic reference frame) of all edge-on
\footnote{We consider a spiral galaxy edge-on if the ratio of semiminor to
semimajor axis is less than $0.175$.}
nearby spiral galaxies in the Principal Galaxy Catalog (PGC) with recession
velocities within various limits. Since we are only interested in the direction
of the rotation axis relative to the SGP, we have reduced all position angles to
between $0^{\circ}$ and $90^{\circ}$.  As shown in Figure~\ref{figs:fig2}a,
there is a clear excess of nearby edge-on galaxies highly inclined relative to
the SGP. The significance of the excess decreases the larger the volume
considered around the Milky Way. Within $1200$ km s$^{-1}$ there are $196$
spirals in the PGC with measured inclinations and position angles. Of these
$\sim 30$ are edge-on. Figure~\ref{figs:fig2}a shows that roughly three times as
many edge-on galaxies with position angles between $0^{\circ}$ and $20^{\circ}$
as in the range ($70^{\circ}$, $90^{\circ}$); a simple KS test shows that this
excess is significant to the $92\%$ level. The magnitude of the excess (and its
significance) decreases to $\sim 40\%$ for spirals with recession velocities
under $2400$ km s$^{-1}$, and is essentially negligible when galaxies within
$5000$ km s$^{-1}$ (or greater) are considered.

This result supports the tendency of spirals to be highly inclined relative to
the SGP claimed by Flin \& Godlowski (1989). Our study provides a compelling
physical interpretation for the origin of such alignment and supports the view
that the angular momentum of spiral galaxies originates in tidal torques tightly
coupled to anisotropies during the expansion, turnaround, and collapse of
protogalactic material.

\section{Summary and Discussion}
\label{sec:disc}

We have used cosmological N-body/gasdynamical simulations to study the
orientation of disk galaxies relative to their surrounding structure. We find
that the angular momentum axis of simulated galaxies is perpendicular to the
minor axis (and very well aligned with the {\it intermediate} axis) of the
material destined to form the galaxy at turnaround, which is usually arranged on
a two-dimensional slab criss-crossed by filaments.  These sheet-like structures
may extend out to several Mpc, in a way reminiscent of the ``supergalactic
plane'' that surrounds the Milky Way. This coherence over a wide range of scales
in the anisotropies present during the expansion and turnaround of protogalactic
material generally results in galaxy disks highly inclined relative to the
large-scale two-dimensional structure where they are embedded.

This provides a natural explanation for the high inclination of the Milky Way
relative to the supergalactic plane, as well as for the excess of nearby edge-on
spirals whose rotation axes lie approximately {\it on} the supergalactic
plane. It may also offer a physical interpretation for the tendency of satellite
galaxies to align along the minor axis of bright spirals (the ``Holmberg
effect''). Our interpretation offers as well a number of ``natural'' predictions
that may be used to falsify it. In particular, we expect an excess of highly
inclined spirals in {\it any} two-dimensional large-scale distribution of
galaxies, a result that may have already been detected in the Perseus
supercluster (Flin 1988), and that could be verified in large galaxy surveys
such as the 2dfGRS and the SDSS. 
The detection of these and other non-trivial correlations between the
spin and matter fields would serve to establish beyond doubt the
validity of tidal torque theory as the origin of the angular momentum
of spiral galaxies.

\acknowledgements 
We thank D.Garc\'\i a Lambas and L.Sales for useful
discussions, and the referee, Avishai Dekel, for a constructive
report. 


\begin{thebibliography}{}

\bibitem[Abadi, Navarro, Steinmetz, \& Eke(2003)]{2003ApJ...591..499A} 
Abadi, M.~G. et al \ 2003, \apj, 
591, 499

\bibitem[Abadi, Navarro, Steinmetz, \& Eke(2003)]{2003ApJ...597...21A} 
Abadi, M.~G. et al \ 2003, \apj, 
597, 21 

\bibitem[Barnes \& Efstathiou(1987)]{1987ApJ...319..575B} Barnes, J.~\& 
Efstathiou, G.\ 1987, \apj, 319, 575 

\bibitem[Dekel(1985)]{1985ApJ...298..461D} Dekel, A.\ 1985, \apj, 298, 461

\bibitem[Doroshkevich (1970)]{390}Doroshkevich A. G., 1970, Astrofiz., 6 581

\bibitem[Flin \& Godlowski(1989)]{1989acfp.proc..418F} Flin, P.~\& 
Godlowski, W.\ 1989, ASSL Vol.~155: Astronomy, Cosmology and Fundamental
Physics, 418


\bibitem[Hoyle (1949)]{396} Hoyle, F., 1949, in Problems of Cosmical Aerodynamics
(Dayton, Ohio: Central Air Documents Office), p.195.
\apjl, 532, L5 

\bibitem[Lee \& Pen(2000)]{2000ApJ...532L...5L} Lee, J.~\& Pen, U.\ 2000, 
\apjl, 532, L5 

\bibitem[Lee \& Pen(2001)]{2001ApJ...555..106L} Lee, J.~\& Pen, U.\ 2001, 
\apj, 555, 106 

\bibitem[Lee \& Pen(2002)]{2002ApJ...567L.111L} Lee, J.~\& Pen, U.\ 2002, 
\apjl, 567, L111 


\bibitem[Meza, Navarro, Steinmetz, \& Eke(2003)]{2003ApJ...590..619M} 
Meza, A. et al\ 2003, \apj, 590, 619 



\bibitem[Paturel et al.(1997)]{1997A&AS..124..109P} Paturel, G., et al.\ 
1997, \aaps, 124, 109 

\bibitem[Peebles(1969)]{1969ApJ...155..393P} Peebles, P.~J.~E.\ 1969, \apj, 
155, 393 

\bibitem[Pen, Lee, \& Seljak(2000)]{2000ApJ...543L.107P} Pen, U., Lee, J., 
\& Seljak, U.\ 2000, \apjl, 543, L107 

\bibitem[Porciani, Dekel, \& Hoffman(2002)]{2002MNRAS.332..325P} Porciani, 
C., Dekel, A., \& Hoffman, Y.\ 2002, \mnras, 332, 325 

\bibitem[Porciani, Dekel, \& Hoffman(2002)]{2002MNRAS.332..339P} Porciani, 
C., Dekel, A., \& Hoffman, Y.\ 2002, \mnras, 332, 339 


\bibitem[Str{\" o}mberg(1934)]{1934ApJ....79..460S} Str{\" o}mberg, G.\ 
1934, \apj, 79, 460 

\bibitem[Sugerman, Summers, \& Kamionkowski(2000)]{2000MNRAS.311..762S} 
Sugerman, B. et al\ 2000, \mnras, 311, 762 

\bibitem[Thacker \& Couchman(2001)]{2001ApJ...555L..17T} Thacker, R.~J.~\& 
Couchman, H.~M.~P.\ 2001, \apjl, 555, L17 

\bibitem[White(1984)]{1984ApJ...286...38W} White, S.~D.~M.\ 1984, \apj, 
286, 38 

\end{thebibliography}
\end{document}